\documentstyle[pre,aps]{revtex}
\begin{document}
%\narrowtext
\draft
\title{L{\'e}vy Flights in External Force Fields: Langevin and Fractional
Fokker--Planck Equations, and their Solutions}
\author{Sune Jespersen\cite{sune}}
\address{Institute of Physics and Astronomy, University of {\AA}rhus,
8000 {\AA}rhus C, Denmark}
\author{Ralf Metzler\cite{ralf}}
\address{School of Chemistry, Tel Aviv University, 69978 Tel Aviv, Israel}
\author{Hans C. Fogedby\cite{hans}}
\address{Institute of Physics and Astronomy, University of {\AA}rhus,
8000 {\AA}rhus C, Denmark,\\
and NORDITA, Blegdamsvej 17, 2100 K{\o}benhavn \O, Denmark}
\date{\today}
\maketitle
\begin{abstract}
We consider L{\'e}vy flights subject to external force fields.
This anomalous transport process is described by two approaches,
a Langevin equation with L{\'e}vy noise and the corresponding
generalized Fokker--Planck equation containing a fractional
derivative in space. The cases of free flights, constant force and
linear Hookean force are analyzed in detail, and we corroborate our
findings with results from numerical simulations. We discuss the
non--Gibbsian character of the stationary solution for the case
of the Hookean force, i.e. the deviation from Boltzmann equilibrium
for long times. The possible connection to Tsallis's $q$--statistics
is studied.
\end{abstract}


\section{Introduction}

In recent years there has been growing interest in anomalous
diffusion in various fields of physics and related sciences. In
one dimension anomalous diffusion is characterized by a mean square
displacement of the form
\begin{equation}
\label{msd}
\langle (\Delta x)^2 \rangle \propto 2D t^{\gamma},
\end{equation}
deviating from the linear dependence on time found for Brownian motion
\cite{blumen,havlin,bouchaud}. The generalized diffusion constant has
the dimension $[D]={\rm cm}^2{\rm sec}^{-\gamma}$.

Subdiffusive transport ($0<\gamma <1$) is encountered in a diversity
of systems, including the charge carrier transport in amorphous
semiconductors \cite{scherlax,schiff}, N.M.R. diffusometry on percolation
structures \cite{kimmich}, and the motion of a bead in a polymer network
\cite{amblard}. On fractal structures in general, subdiffusion prevails
due to the occurrence of holes of all length scales \cite{havlin}. Examples
of enhanced diffusion ($\gamma>1$) include tracer particles in vortex
arrays in a rotating flow \cite{weeks}, layered velocity fields
\cite{zuklablu}, and Richardson diffusion \cite{richardson}.

L{\'e}vy flights are used to model a variety of processes, such as
bulk mediated surface diffusion \cite{bychuk} and applications in
porous glasses and eye lenses \cite{kimmich1}, transport in micelle
systems or heterogeneous rocks \cite{rocks}, special problems in reaction
dynamics \cite{react}, Single Molecule Spectroscopy \cite{sms}, and even
the flight of an albatross \cite{albatros}.

Among the different frameworks for describing anomalous diffusion
are fractional Brownian motion \cite{mandelbrot}, the continuous
time random walk scheme \cite{scherlax,klablushle}, fractional
diffusion equations \cite{schneider,fde}, generalized Langevin and
Fokker--Planck equations (FPEs)
\cite{fogedby1,fogedby,zaslavsky,kolwankar,prl}, and generalized
thermostatistics \cite{tsallis,alemanyi}.
Common to all these approaches is the violation of the Central Limit
Theorem of probability theory \cite{levy,gnedenko}, and this is achieved
either by correlations or by long--tailed statistics. L{\'e}vy statistics
\cite{levy,gnedenko} is an example of the latter, and has been used
extensively to model both enhanced and dispersive diffusion
\cite{bouchaud,klablushle,levytopics}. The two most fundamental properties
of the L{\'e}vy
distributions are the stability under addition, following from the
Generalized Central Limit Theorem valid for L{\'e}vy distributions,
and the asymptotic power law decay. These features are responsible
for the anomalous character of the diffusion processes we have in mind.

A fractional Fokker--Planck equation (FFPE) describing anomalous
transport close to thermal equilibrium, was presented recently
\cite{prl}. Since it describes subdiffusion in the force--free case, it
involves a strong, i.e. slowly decaying memory. In the present paper, we
focus
on FFPEs which are connected with L{\'e}vy flights, and are based
on the following Langevin equation for the coordinate $x(t)$
\cite{fogedby1,fogedby}:
\begin{equation}
\label{langevin1}
\frac{d}{dt}x(t)= \frac{F(x)}{\gamma m}+\eta(t).
\end{equation}
Here, $m$ is the mass of the diffusing particle, and $\gamma$ denotes
a friction coefficient. $F(x)$ is the external force field.
For simplicity we shall work in one dimension, with obvious modifications
in the general case. The noise $\eta(t)$ is the source of the anomalous
behavior. We assume $\eta(t)$ to be uncorrelated at different times, and
to obey L{\'e}vy statistics \cite{rem2}. In Fourier space we thus define
\begin{equation}
p(k)= \int d \eta e^{-ik\eta} p(\eta)=\exp (-D|k|^\mu),
\label{levy}
\end{equation}
where $0<\mu<2$. The probability density function (PDF) $p(\eta)$ has
an asymptotic power--law behavior according to $p(\eta)\sim|\eta|^{-1-\mu}$
\cite{levy,gnedenko,rem1}. Here and in the following, we denote the Fourier
transform
of a function by using the explicit dependence on the wave number $k$, and
analogously $u$ for the Laplace transform of a $t$--dependent function.
For the special case $\mu=2$ in Eq. (\ref{levy}), i.e. for
a Gaussian noise, we are led back to the Brownian case. In this case $2D$
is
the variance of the PDF, but even in the general case the parameter $D$
characterizes the width of the PDF in some sense.

The Langevin equation Eq. (\ref{langevin1}) is a stochastic differential
equation. Often it is more convenient to work with the deterministic
equation for the distribution function, the FPE \cite{risken,caffey}.
For the power--law noise $\eta(t)$ defined through Eq. (\ref{levy}),
we are led to the following FFPE \cite{fogedby1,fogedby}:
\begin{equation}
\label{fp1}
\frac{\partial}{\partial t}W(x,t)=-\frac{\partial}{\partial x}\left(\frac{
F(x)W(x,t)}{\gamma m}\right)+ D\nabla^\mu W(x,t).
\end{equation}
Here, $D$ denotes the generalized diffusion coefficient with the
dimension $[D]={\rm cm}^{\mu}{\rm sec}^{-1}$.
The Riesz fractional derivative in Eq. (\ref{fp1}) is defined through its
Fourier--transform \cite{oldham,samko}
\begin{equation}
\label{riesz}
\nabla^\mu=-\int \frac{d^dk}{(2\pi)^d}e^{i{\mathbf k}\cdot{\mathbf
x}} |k|^\mu
\end{equation}
in $d$ dimensions. Note that in the FFPE Eq. (\ref{fp1}) the first order
differential operator acting upon the force term is not affected by
the introduction of the L{\'e}vy distribution Eq. (\ref{levy}), see Ref.
\cite{fpder} where a unifying derivation of FFPEs from a generalized
master equation is discussed.

In the following we will consider the FFPE Eq. (\ref{fp1}) for the three
cases of the free flight ($F=0$), the constant force $F(x)=F_0$, and
the Hookean force $F(x)=-\lambda x$, comparing to the Brownian case
as we go along. We will discuss the differences to the subdiffusive
FFPE of Ref. \cite{prl}, where a fractional operator in time is
encountered and the spatial part of the standard FPE remains unchanged,
as well as the possible connection to Tsallis's $q$--statistics.
Numerical simulations corroborate our theoretical findings. We then
exemplify the method of solution for the Langevin equation
Eq. (\ref{langevin1}) for a linear force with an additional drift term.
Before drawing the conclusions, we give some remarks on the simulations.
Some additional calculations on the nature of the correlation functions
are presented in the Appendix.

\section{Free L{\'e}vy flight}

\begin{figure}
\unitlength=1cm
\begin{center}
\begin{picture}(6,4.8)
\put(-2.4,-5.2){
\includegraphics{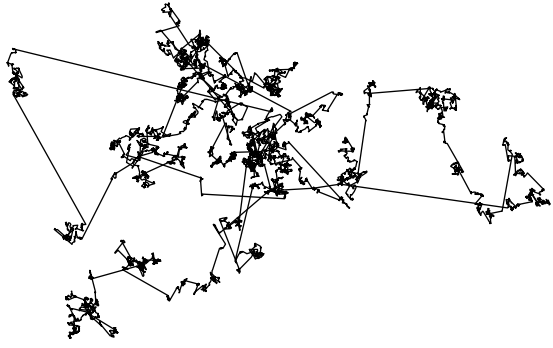}}
\end{picture}
\end{center}
\caption{
Typical L{\'e}vy--flight for the L{\'e}vy index $\mu=1.4$. The clustering
is obvious. Each cluster is statistically self--similar to the unmagnified
picture. The fractal dimension of the flight is $d_f=\mu$
\protect\cite{bouchaud,fogedby1}.
\label{fig1}}
\end{figure}

In this case we have to solve the anomalous diffusion equation
\begin{equation}
\label{free}
\frac{\partial}{\partial t}W(x,t)=D\nabla^\mu W(x,t).
\end{equation}
Fourier transforming Eq. (\ref{free}) and utilizing the definition of the
fractional Riesz operator Eq. (\ref{riesz}) we have
\begin{equation}
\label{free_fourier}
\frac{\partial}{\partial t}W(k,t)=-D|k|^\mu W(k,t)
\end{equation}
with the solution
\begin{equation}
\label{solution1}
W(k,t)=e^{-Dt|k|^\mu},
\end{equation}
demanding the sharp initial condition $x(0)=0$, corresponding to
$W(x,0)=\delta(x)$ or $W(k,0)=1$.
Comparing to Eq. (\ref{levy}) we recognize the characteristic function of
the
L{\'e}vy distribution, and we thus find in real space the stable law
$L_{\mu}$:
\begin{eqnarray}
\label{x-space}
W(x,t)&=& (Dt)^{-1/\mu} L_\mu
\left(\frac{|x|}{[Dt]^{1/\mu}}\right)\nonumber\\
&=&\frac{\pi}{\mu|x|}H^{1,1}_{2,2}\left[\frac{|x|}{(Dt)^{1/\mu}}\left|
\begin{array}{l} (1,1/\mu),(1,1/2)\\ (1,1),(1,1/2)
\end{array} \right. \right].
\end{eqnarray}
In Eq. (\ref{x-space}), we have expressed the L{\'e}vy distribution
exactly in terms of Fox' $H$--functions \cite{mathai,west}. This result
Eq. (\ref{x-space}) is expected, due to the stable
law nature of the underlying L{\'e}vy distribution. The asymptotic behavior
of  the propagator $W(x,t)$ can be derived from Eq. (\ref{x-space})
and reads
\begin{equation}
\label{levas}
W(x,t)\sim \frac{Dt}{|x|^{1+\mu}}
\end{equation}
for $|x|^{\mu}/[Dt]\gg 1$, and thus we encounter a divergence of the mean
square displacement at all times: $\langle x^2(t)\rangle=\infty$. This is
intuitively clear due to the occurrence of arbitrarily long jumps in
the L{\'e}vy flight, see Fig. \ref{fig1}.
%\begin{figure}
%\unitlength=1cm
%\begin{center}
%\begin{picture}(6,4.8)
%\put(-2.4,-5.2){
%\special{psfile=fig1.ps hoffset=0
%          voffset=0 hscale=48 vscale=48 angle=0}}
%\end{picture}
%\end{center}
%\caption{
%Typical L{\'e}vy--flight for the L{\'e}vy index $\mu=1.4$. The clustering
%is obvious. Each cluster is statistically self--similar to the unmagnified
%picture. The fractal dimension of the flight is $d_f=\mu$
%\protect\cite{bouchaud,fogedby1}.
%\label{fig1}}
%\end{figure}
Mathematically, the divergence is evident from Eq. (\ref{solution1})
by using the properties of the characteristic function: $\langle
x^n(t) \rangle=\left.i^n d^nW(k,t)/dk^n\right|_{k=0}$.

\begin{figure}
\unitlength=1cm
\begin{center}
\begin{picture}(6,5.8)
\put(-2.2,6.6){
\includegraphics{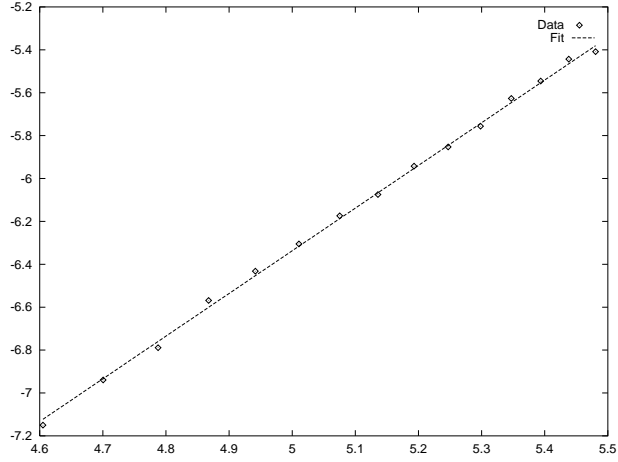}}
\end{picture}
\end{center}
\caption{Function $\langle x^2(t)\rangle_L$ from Eq.
(\protect\ref{meansquare}) versus time with $\mu=1$ in a $\log$--$
\log$ plot. The slope of the straight line is $1.990\pm 0.028$,
which is to be compared to the expected value $2/\mu=2$.
\label{fig2}}
\end{figure}

\begin{figure}
\unitlength=1cm
\begin{center}
\begin{picture}(6,5.8)
\put(-2.2,6.6){
\includegraphics{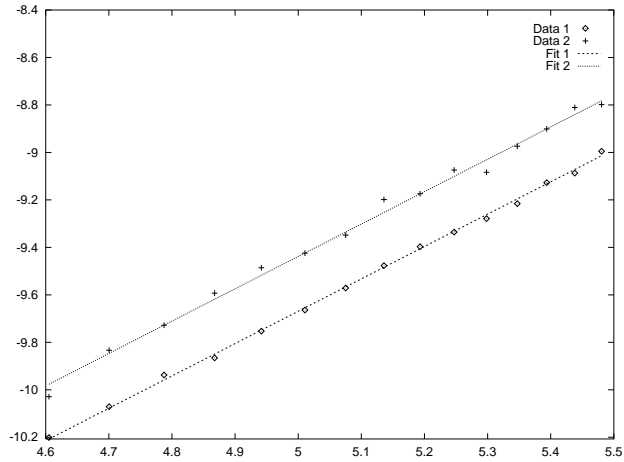}}
\end{picture}
\end{center}
\caption{$\log$--$\log$ plot of $\langle x^2 \rangle_L$ versus
time from Eq. (\protect\ref{meansquare}) (lower curve), compared to the
squared absolute mean $\langle|x|\rangle^2$ from Eq. (\protect\ref{sam})
(upper curve) for $\mu=1.5$. The fitted slope is $1.363 \pm 0.013$ and
$1.364 \pm 0.035$, respectively, which is in good agreement with the
theoretical value $4/3$.
\label{fig3}}
\end{figure}

\begin{figure}
\unitlength=1cm
\begin{center}
\begin{picture}(6,6.2)
\put(-2.2,6.6){
\includegraphics{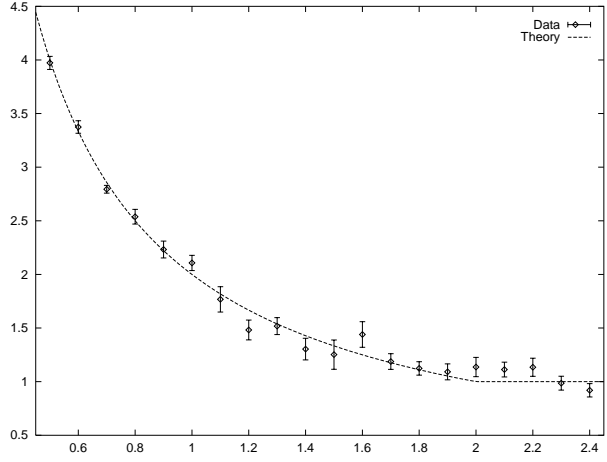}}
\end{picture}
\end{center}
\caption{Graph of the slope $2/\mu$ according to Eq.
(\protect\ref{meansquare}) as a function of the L{\'e}vy index $\mu$.
Note the bend at $\mu=2$ marking the transition to normal diffusion.
\label{fig4}}
\end{figure}

In order to extract the scaling form implied by Eq. (\ref{x-space})
operationally, one could enclose the ``walker'' in an imaginary growing box
(see Sec. \ref{simulations}):
\begin{equation}
\label{meansquare}
\langle x^2(t)\rangle_L\sim\int_{L_1t^{1/\mu}}^{L_2t^{1/\mu}}dx\,
x^2W(x,t)\sim t^{2/\mu}\;.
\end{equation}
This has been implemented numerically, and as can be seen from Fig.
\ref{fig2}, where we have a straight line on a log--log plot of $\langle
x^2\rangle_L$ as a function of $t$, for a fixed $\mu$, the expected
power law index $2/\mu$ according to Eq. (\ref{meansquare}) is
found.
%\begin{figure}
%\unitlength=1cm
%\begin{center}
%\begin{picture}(6,5.8)
%\put(-2.2,6.6){
%\special{psfile=fig2.ps hoffset=0
%          voffset=0 hscale=34 vscale=34 angle=270}}
%\end{picture}
%\end{center}
%\caption{Function $\langle x^2(t)\rangle_L$ from Eq.
%(\protect\ref{meansquare}) versus time with $\mu=1$ in a $\log$--$
%\log$ plot. The slope of the straight line is $1.990\pm 0.028$,
%which is to be compared to the expected value $2/\mu=2$.
%\label{fig2}}
%\end{figure}
This scaling result $\langle x^2 (t) \rangle_L \sim t^{ 2/\mu}$ is not to
be
confused with the mean square displacement $\langle x^2 \rangle=\int dx \,
x^2W(x,t)=\infty$ for the L{\'e}vy flight, resulting from Eq.
(\ref{levas}).
However, for $\mu>1$, the squared absolute mean
\begin{equation}
\label{sam}
\langle |x| \rangle^2=\left(\int d x \, |x|W(x,t)\right)^2
\end{equation}
converges, and is proportional to $\langle x^2 (t) \rangle_L$ from Eq.
(\ref{meansquare}), see the discussion in Ref. \cite{reply}. Fig.
\ref{fig3} shows this proportionality for $\mu=1.5$ \cite{comment}.
%\begin{figure}
%\unitlength=1cm
%\begin{center}
%\begin{picture}(6,5.8)
%\put(-2.2,6.6){
%\special{psfile=fig3.ps hoffset=0
%          voffset=0 hscale=34 vscale=34 angle=270}}
%\end{picture}
%\end{center}
%\caption{$\log$--$\log$ plot of $\langle x^2 \rangle_L$ versus
%time from Eq. (\protect\ref{meansquare}) (lower curve), compared to the
%squared absolute mean $\langle|x|\rangle^2$ from Eq. (\protect\ref{sam})
%(upper curve) for $\mu=1.5$. The fitted slope is $1.363 \pm 0.013$ and
%$1.364 \pm 0.035$, respectively, which is in good agreement with the
%theoretical value $4/3$.
%\label{fig3}}
%\end{figure}
In Fig. \ref{fig4}, we graph the slope $2/\mu$ as a function of $\mu$ for
a variety of values for the L{\'e}vy index $\mu$, and we obtain excellent
agreement with Eq. (\ref{meansquare}).
%\begin{figure}
%\unitlength=1cm
%\begin{center}
%\begin{picture}(6,6.2)
%\put(-2.2,6.6){
%\special{psfile=fig4.ps hoffset=0
%          voffset=0 hscale=34 vscale=34 angle=270}}
%\end{picture}
%\end{center}
%\caption{Graph of the slope $2/\mu$ according to Eq.
%(\protect\ref{meansquare}) as a function of the L{\'e}vy index $\mu$.
%Note the bend at $\mu=2$ marking the transition to normal diffusion.
%\label{fig4}}
%\end{figure}
In the case $\mu=2$ we see from Eq. (\ref{solution1}) and the results of
the
simulations in Fig. \ref{fig4} that the usual Brownian behavior is
recovered.
Especially, we obtain the mean square displacement
\begin{equation}
\label{meansquare2}
\langle x^2(t)\rangle=2Dt.
\end{equation}
The properties of free L{\'e}vy flights
could also be obtained directly from Eq. (\ref{langevin1}) employing the
method of characteristic functions. This view also allows us to extract the
distribution of speeds, which turns out to be L{\'e}vy
distributed. From this likewise follows the mean kinetic energy, and we
have for a finite mass $m$ of the walker in the case $\mu<2$
\begin{equation}
\label{kinetic}
\langle\frac{1}{2}mv^2\rangle=\infty.
\end{equation}

\section{Constant force: Drift and Acceleration}

For a constant force $F(x)=F_0$, the FFPE Eq. (\ref{fp1}) reads
\begin{equation}
\label{constforce}
\frac{\partial}{\partial t}W(x,t)=-\frac{\partial}{\partial x}\left(
\frac{F_0W(x,t)}{\gamma m}\right)+D\nabla^\mu W(x,t).
\end{equation}
Returning to the Fourier domain, we recover the equation
\begin{equation}
\label{constforce_fourier}
\frac{\partial}{\partial t}W(k,t)=\left(-ik\frac{F_0}{\gamma m}
-D|k|^\mu\right)W(k,t),
\end{equation}
which for the propagator, i.e. $W(k,t)$ with the initial condition
$W(k,0)=1$, yields
\begin{equation}
\label{sol2_fourier}
W(k,t)=\exp \left(-t\left[ik\frac{F_0}{\gamma m}+D|k|^\mu \right]\right).
\end{equation}
This is the same L{\'e}vy distribution as calculated for the free
L{\'e}vy flight in Eq. (\ref{solution1}), but at the translated
coordinate
\begin{equation}
\label{solution2}
W(x,t)=W_0\left(x-\frac{F_0t}{\gamma m},t\right).
\end{equation}
Here $W_0$ refers to the distribution of the free
L{\'e}vy flight. The displacement of the coordinate is due to the
balancing of the friction against the imposed constant force,
i.e.  $\gamma m v=F_0$, in the Galilei transformed system $x\rightarrow
x-F_0t /[\gamma m]$. Clearly, the analytical form for the solution
in $x$--space is still given by Eq.
(\ref{x-space}), but now for the translated coordinate. Thus L{\'e}vy
flights in a constant force field described by Eq. (\ref{constforce}) are
similar to (anomalous) diffusion in a constant velocity field
\cite{pre1,levy1}. The
reason why we go directly to the steady state described by $\gamma m v
=F_0$ is due to the omission of the inertial term in the Langevin
equation Eq. (\ref{langevin1}). By including this term it can be shown
that we obtain a transient contribution. Thus the diffusion in a force
field is only equivalent to diffusion in a constant velocity field
for large times, according to the discussion in Ref. \cite{pre1}, see
below. In fact the same situation is encountered even for the standard
diffusion--advection picture \cite{pre1,levy1,vankampen}. For long waiting
periods in the random walk picture for the subdiffusive model,
the external force can only act upon the walker, when it is released from
a trap, which leads to the sublinear dependence $\langle x(t) \rangle
\propto t^{\gamma}$ found in Ref. \cite{pre1}, where $\gamma$
denotes the power law index of the broad waiting time distribution.
The Poissonian waiting time distribution, on the other hand, which
is typical for the L{\'e}vy flyer, leads to the same behavior
Eq. (\ref{mean}) below [for $1<\mu\le2$], as known from the Brownian
case, due to the very short waiting periods in between jumps. The
long--tailed nature of the L\'evy flight however makes all higher moments
infinite.
Both effects lead to an accelerated time dependence of the motion.
Including the inertial term in the Langevin equation
Eq. (\ref{langevin1}) changes the solution for short times according
to
\begin{eqnarray}
\nonumber
x(t)&=&\frac{F_0}{m\gamma} \left(t-\frac{1-e^{-\gamma t}}{
\gamma}\right)\\
&&+\int_0^t ds \, \eta(s) \left(1-e^{-\gamma
(t-s)}\right)\,.
\label{inert}
\end{eqnarray}
At times much greater than the characteristic time $\gamma^{-1}$,
we have
\begin{equation}
x(t) \simeq \frac{F_0}{m\gamma}t+\int_0^t ds \eta(s)\,,
\end{equation}
which follows from the behavior of the Laplace transform
of the convolution integral in the second summand of Eq. (\ref{inert}):
$\frac{\eta(u)}{u} \left(1-\frac{u}{u+\gamma}\right)\sim\frac{\eta(u)}{u}$
in the limit $u\ll\gamma$. Thus the effect of the inertial term is
negligible for times $t$ much greater than $\gamma^{-1}$.

If the first moment exists, that is for $1<\mu\leq 2$, we find
from Eq. (\ref{sol2_fourier})
\begin{equation}
\label{mean}
\langle x(t)\rangle= \frac{F_0t}{\gamma m}\,.
\end{equation}
Only for $\mu=2$ we have a finite second moment, and the mean square
displacement becomes $\langle \left[(x(t)-\langle x(t)\rangle\right]
^2\rangle=2Dt$, in agreement with Eqs. (\ref{meansquare2}) and
(\ref{solution2}).

For the standard FPE as well as for the subdiffusive FFPE introduced
in \cite{prl}, one finds the generalized Einstein relation
\cite{prl,barkai,vankampen} $\langle x(t)\rangle_{F_0}=F_0\langle x^2(t)
\rangle_0/(2k_BT)$, relating the first moment in presence of the force
$F_0$ to the second moment in absence of the force. For the L{\'e}vy
flight model defined in Eq. (\ref{constforce}), only in the Brownian
limit $\mu=2$ this relation is satisfied, provided we choose
the proper amplitude of the noise, i.e. take it to be thermal noise:
$D=k_BT/[\gamma m]$. Generally since we have a diverging mean square
displacement, the generalized Einstein relation does not hold, and we have
a violation of the classical fluctuation--dissipation theorem.

\section{Linear Force and Non--Gibbsian stationary solution}

In the case of ordinary Brownian motion the diffusing particles can be
trapped in a harmonic potential and thus attain an equilibrium
distribution with a finite variance \cite{vankampen,risken}. More precisely
this equilibrium distribution is the Gibbs or Boltzmann
distribution, also obtainable from maximizing the Gibbs entropy under
the constraints of norm and energy conservation. Also for
subdiffusive transport in a harmonic potential, this property
is fulfilled \cite{prl}. For L{\'e}vy flights we shall see that a
stationary solution does exist, however it possesses
{\it no\/} finite variance. This deviation from the Gibbs--Boltzmann
equilibrium implies that L{\'e}vy flights do not describe systems
close to thermal equilibrium.

For the Hookean force $F(x)=-\lambda x$, corresponding to the
harmonic potential $V(x)=\frac{1}{2}\lambda x^2$, the FFPE
Eq. (\ref{fp1}) becomes
\begin{equation}
\label{linear}
\frac{\partial}{\partial t}W(x,t)=\frac{\partial}{\partial x}\left(
\frac{\lambda}{\gamma m}x W(x,t)\right)+D\nabla^\mu W(x,t).
\end{equation}
In Fourier--space, the conjugate equation reads
\begin{equation}
\label{linear_fourier}
\frac{\partial}{\partial t}W(k,t)=-\frac{\lambda}{\gamma m}k\frac{
\partial}{\partial k} W(k,t)-D|k|^\mu W(k,t),
\end{equation}
which can be easily solved by making a transformation of variables
(applying the method of characteristics)
\begin{equation}
\label{lin_four_sol}
W(k,t)=\exp\left(-\frac{\gamma m D|k|^\mu}{\mu\lambda}\left[1-e^{-
\mu\lambda t/ \gamma m}\right]\right).
\end{equation}
This is still a L{\'e}vy distribution, only with a different ``width''
$D\rightarrow \frac{\gamma m D}{\mu\lambda}(1-e^{-\mu\lambda t/ \gamma
m})$, and the exact solution in real space can again be obtained from
Eq. (\ref{x-space}) by inserting the time--dependent width.
For $\mu=2$ we recover the Brownian results, but in the general
case $\mu<2$ a different situation arises. We always reach a stationary
distribution
\begin{equation}
\label{stationary}
W_{\rm st}(k)=\exp\left(-\frac{\gamma m D|k|^\mu}{\mu\lambda}\right),
\end{equation}
but with a diverging mean square. The exact stationary solution in
$x$--space
can be given in terms of Fox's $H$--functions:
\begin{equation}
\label{harmstat}
W_{\rm st}(x)=\frac{\pi}{|x|}
H^{1,1}_{2,2} \left[ \frac{|x|^{\mu}\mu \lambda}{D\gamma m}
\left| \begin{array}{l} (1,1), \left(1,\frac{\mu}{2}\right)\\
(1,\mu),\left(1,\frac{\mu}{2}\right) \end{array}
\right. \right],
\end{equation}
leading to the asymptotic power--law behavior $W_{\rm st}(x) \sim D
\gamma m/(\mu \lambda |x|^{1+\mu})$.

\begin{figure}
\unitlength=1cm
\begin{center}
\begin{picture}(6,6.2)
\put(-2.2,6.6){
\includegraphics{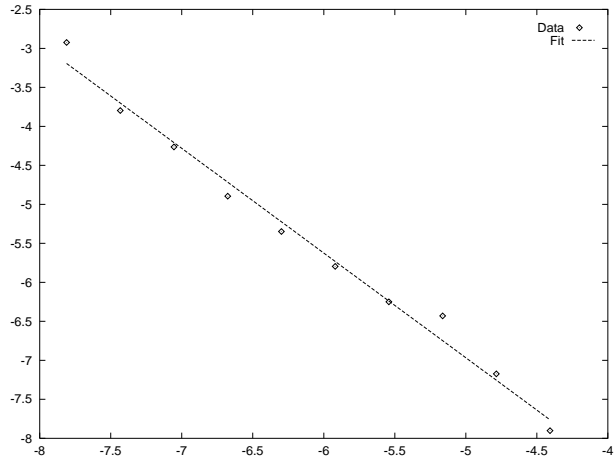}}
\end{picture}
\end{center}
\caption{Histogram for the stationary solution $W_{\rm st}$ of a L{\'e}vy
flight in a harmonic potential versus $|x|$, as a plot of $\log W_{\rm
  st}(y)$ versus$ y $, where $y=\log(x)$ denotes the natural logarithm of
the
position of the flyer, see Sec. \protect\ref{simulations}. The data were
produced for a L{\'e}vy index $\mu=1.4$. The fit indicated by the dashed
line reveals a slope of $-1.408 \pm 0.108$, which thus shows a good
agreement
with the theoretical prediction $-\mu$.
\label{fig5}}
\end{figure}

A numerical result for a simulation of a L{\'e}vy flight in an
harmonic potential is shown in Fig. \ref{fig5}. The slope is in
good agreement with the theoretical prediction.
%\begin{figure}
%\unitlength=1cm
%\begin{center}
%\begin{picture}(6,6.2)
%\put(-2.2,6.6){
%\special{psfile=fig5.ps hoffset=0
%          voffset=0 hscale=34 vscale=34 angle=270}}
%\end{picture}
%\end{center}
%\caption{Histogram for the stationary solution $W_{\rm st}$ of a L{\'e}vy
%flight in a harmonic potential versus $|x|$, as a plot of $\log W_{\rm
%  st}(y)$ versus$ y $, where $y=\log(x)$ denotes the natural logarithm of
%the
%position of the flyer, see Sec. \protect\ref{simulations}. The data were
%produced for a L{\'e}vy index $\mu=1.4$. The fit indicated by the dashed
%line reveals a slope of $-1.408 \pm 0.108$, which thus shows a good
%agreement
%with the theoretical prediction $-\mu$.
%\label{fig5}}
%\end{figure}

We pause to mention that we could have derived the solution
Eq. (\ref{lin_four_sol}) also by means of a separation ansatz, i.e.
assuming
a particular solution of the form $W_n(x,t)=T_n(t)\varphi_n(x)$, as
was discussed in Refs. \cite{prl,pre1}. Thus we arrive at the ordinary
differential equations
\begin{mathletters}
\begin{eqnarray}
\label{dec1}
&&\frac{d}{dt}T(t)=-\lambda_n T(t)\\
&&\lambda_n\varphi_n(x)+\frac{d}{dx} \left[ \frac{\lambda}{\gamma m}
x \varphi_n(x)\right]+D \nabla^{\mu}\varphi_n(x)=0
\label{dec2}
\end{eqnarray}
\end{mathletters}
\noindent with the eigenvalue $\lambda_n$. The complete solution is then
given
by the sum $W(x,t)=\sum\limits_{n=0}^{\infty}W_n(x,t)$. For the
time behavior we find the usual exponentially decaying modes,
$T_n(t)=e^{-\lambda_nt}$, and for the spatial eigenfunction we have
\begin{equation}
\varphi_n(k)=c_n|k|^{\lambda_n\gamma m/\lambda} e^{-|k|^{\mu}/\mu}.
\end{equation}
The eigenvalues are given by $\lambda_n=\frac{\lambda}{\gamma m}
\mu n$, and the complete solution in wave number space is given by
\begin{eqnarray}
\nonumber
W(k,t)&=&e^{-D\gamma m|k|^{\mu}/[\mu \lambda]}\\
&&\times \sum_{n=0}^{\infty}
\frac{1}{n!} \left( \frac{D\gamma m}{\mu \lambda}\right)^n
|k|^{\mu n} e^{-\mu n \lambda t/[\gamma m]}.
\end{eqnarray}
The sum converges to Eq. (\ref{lin_four_sol}), and
can be transformed back to real space by use of the Fox
functions:
\begin{eqnarray}
\nonumber
W(x,t)&=&\sum_{n=0}^{\infty} \frac{1}{n!} \left( \frac{D\gamma m}{
\mu \lambda}\right)^n \frac{\pi}{|x|} \mu^{\lambda n/[\gamma m]}
e^{-\mu n \lambda t/[\gamma m]}\\
&&\hspace{-0.2in}
\times H^{1,1}_{2,2} \left[\frac{\mu \lambda}{D \gamma m}
|x|^{\mu} \left| \begin{array}{l} \left(1-\frac{\lambda n}{
\gamma m},1\right),\left(1,\frac{\mu}{2}\right)\\
(1,\mu),\left(1,\frac{\mu}{2}\right) \end{array}
\right. \right].
\label{harmsol}
\end{eqnarray}
In comparison to Risken's result for the Fokker--Planck
equation in a harmonic potential in the Brownian case
\cite{risken}, also referred to as the Ornstein--Uhlenbeck
process, the eigenvalues in the solution Eq. (\ref{harmsol}) for
$\mu=2$ take on only even numbers. This
is due to our consideration of the start in the origin, so that
all the uneven Hermite polynomials occurring in the solution
given in Ref. \cite{risken}, vanish: $H_{2n+1}(0)=0$.
We note that the Fox functions in Eq. (\ref{harmsol})
can be considered as generalized Hermite polynomials.
Clearly, the stationary solution corresponding to
$\lambda_0=0$ obtained from Eq. (\ref{harmsol}) is the same
stable law as in Eq. (\ref{harmstat}). Also it can be seen,
that only in the Brownian case $\mu=2$ do we recover the Boltzmann
distribution $W(x)\propto e^{-\lambda x^2/[2k_BT]}$. Thus
Boltzmann equilibrium with a finite variance is
not reached, in spite of the fact that the system is isolated and
time--independent in respect to the ensemble. This can be {\it
  physically\/} understood
as a consequence of the diverging mean kinetic energy of the free
L{\'e}vy flight. In this case we have
\begin{equation}
\label{brown}
\frac{d}{dt}v(t)=-\gamma v(t)+\gamma \eta(t) .
\end{equation}
If we take the noise to be white according to Eq. (\ref{levy}) with
$\mu=2$, we
get from Eq. (\ref{brown}) that in the stationary state
\begin{eqnarray}
\label{kinetic2}
\langle v^2\rangle=D\gamma &\Leftrightarrow&\langle
E_{kin}\rangle=\frac{m\gamma D}{2}.
\end{eqnarray}
When the external harmonic potential is turned on, a {\it length
scale\/} is introduced by the comparison $\langle E_{kin}\rangle
\approx\langle E_{pot}\rangle=\frac{1}{2}\lambda \langle x^2 \rangle$.
This means that $\langle x^2 \rangle\approx mD\gamma/\lambda$. In fact,
solving Eq. (\ref{langevin1}) in the Brownian case with the
harmonic potential, we obtain exactly $\langle x^2 \rangle=mD\gamma
/\lambda$, in accordance with the equipartition theorem. Similar
considerations remain valid for the subdiffusive FFPE
from Ref. \cite{pre1}, which thus describes anomalous systems
close to thermal equilibrium. However, in the L{\'e}vy case we have
\begin{equation}
\label{levykin}
\langle E_{kin}\rangle=\infty,
\end{equation}
and therefore the length scale which appears when the potential
is introduced, is also diverging. Consequently the question arises whether
other statistics could predict the equilibrium
distribution in the present context. We here consider the recently
proposed Tsallis's $q$--statistics \cite{tsallis}, according to which
the generalized entropy
\begin{equation}
\label{entropi}
S_q[p(x,v)]=\frac{1-\int p^q(x,v)\,dxdv}{q-1}
\end{equation}
is introduced along with the generalized constraints
\begin{mathletters}
\begin{eqnarray}
\label{gencon}
\int p(x,v)\,dx&=&1\\
\int p^q(x,v)E(x,v)\,dx&=&U;
\label{gencona}
\end{eqnarray}
\end{mathletters}
for $q\rightarrow 1$, $S_q$ recovers the usual Boltzmann entropy.
Here $v$ is the velocity of the particle and $E(x,v)$ is its energy.
Thus Eq. (\ref{gencona}) is a generalized constraint of
conservation of energy along with the usual norm conservation
Eq. (\ref{gencon}).
Varying Eq. (\ref{entropi}) subject to these constraints by introducing
Lagrange--multipliers, one obtains the stationary
distribution
\begin{equation}
\label{stat}
p_q(x,v)\sim\left(1-(1-q)\frac{\beta}{2} \left[\lambda x^2+mv^2
\right]\right)^{
1/(1-q)}
\end{equation}
Integrating over all velocities to obtain the distribution of positions
alone
yields
\begin{equation}
\label{statpos}
p_q(x)\sim\left(1-(1-q)\frac{\beta}{2} \lambda
x^2\right)^{\frac{3-q}{2-2q}},
\end{equation}
compare to Ref. \cite{alemanyi}.
Matching this expression to the asymptotic behavior of the stationary
solution
of the FFPE, Eq. (\ref{stationary}), we obtain
\begin{equation}
\label{q/u}
\mu=\frac{4-2q}{q-1}
\end{equation}
The relationship Eq. (\ref{q/u}) implies that $q$ can range in the interval
$(1,2)$.
This is at variance with the relation $\mu=(3-q_{\rm free})/(q_{\rm
free}-1)$
found in the case of the free L{\'e}vy flight \cite{alemanyi,levytopics},
where the allowed range is $q_{\rm free} \in (5/3,3)$. Note that
the $q$--$\mu$ relation does not involve the
potential strength $\lambda$. Thus, the Tsallis index $q$ makes a jump
for a given L{\'e}vy index $\mu$, when the potential is switched off,
irrespectively of how slowly, e.g. quasi--statically, the limit $\lambda
\to 0$ is performed. Moreover, away from the asymptotic regime,
the solution predicted by the entropy Eq. (\ref{entropi}) does not agree
with the solution found in Eq. (\ref{stationary}) in the stationary
state. Thus we conclude, that the Tsallis entropy is not the appropriate
framework
for L{\'e}vy flights in a harmonic potential described by the
generalized Fokker--Planck equation Eq. (\ref{fp1}). This form of a
generalized entropy does not give rise to the solution of Eq.
(\ref{stationary}). Recently, it
has also been shown that Tsallis $q$--statistics is by no means
unique \cite{papa}, so that it would have been rather surprising if
this very special form of statistics had led to the complete
description according to the L{\'e}vy flight model. Finally, the
FFPE Eq. (\ref{fp1}) being linear, is not compatible with the
non--extensive
nature of Tsallis entropy \cite{tsallis}; compare with the non--linear
diffusion equation derived from Tsallis entropy in Ref. \cite{compte}.

\begin{figure}
\unitlength=1cm
\begin{center}
\begin{picture}(6,6.2)
\put(-2.2,6.6){
\includegraphics{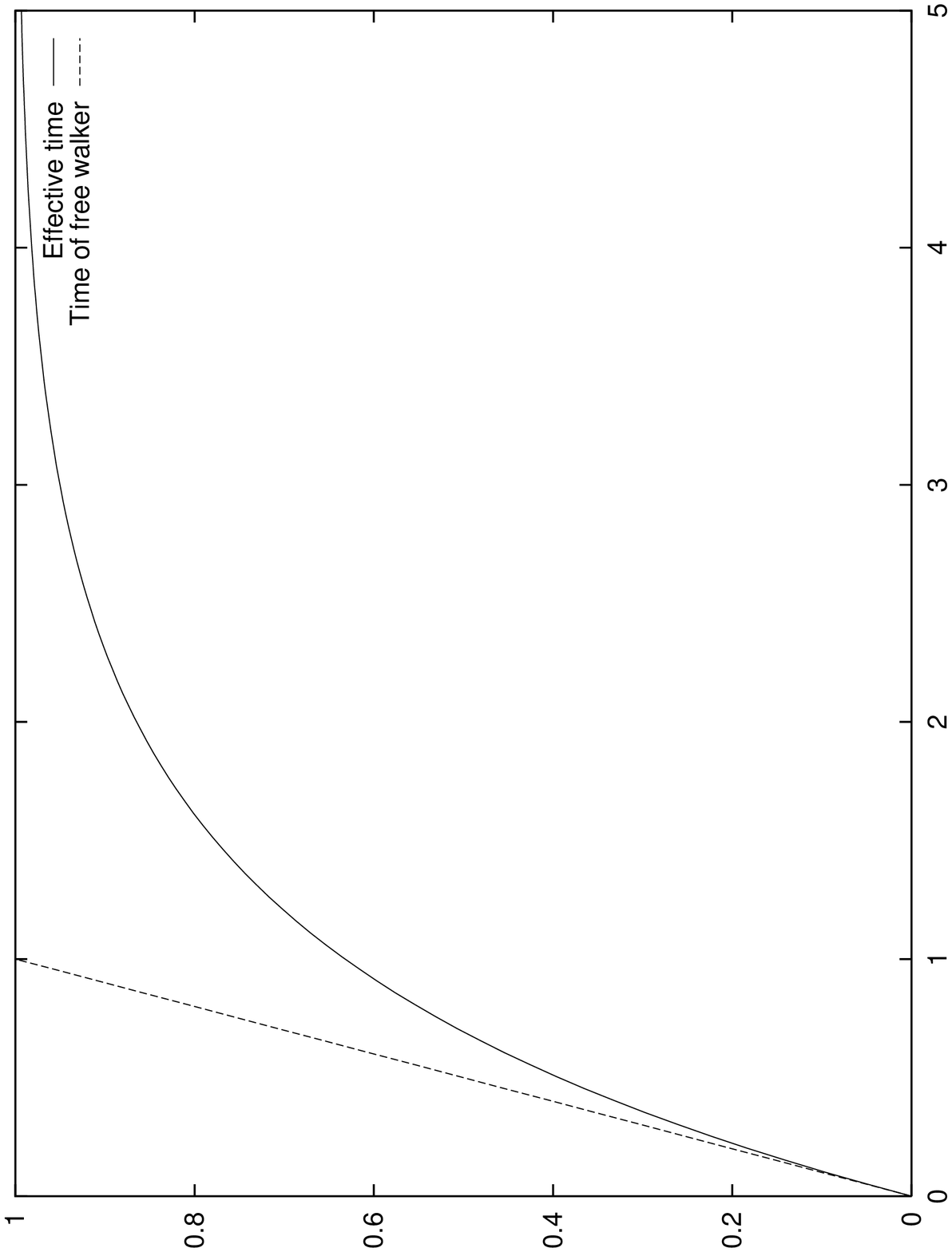}}
\end{picture}
\end{center}
\caption{
The linear time as seen by the free walker (dashed line), compared
to the effective time sensed by the random walker in the harmonic
potential, as a function of the laboratory time. For the L{\'e}vy
flight in the potential, the restoring Hookean force slows down the
spreading of the diffusion, and eventually brings it to a halt.
\label{fig6}}
\end{figure}

By comparing the distribution of the particle in the harmonic potential
$W(x,t)$ with that of the free flight $W_0(x,t)$, we obtain the
correspondence
\begin{equation}
\label{correspondence}
W(x,t)=
W_0\left(x,t_{\mbox{eff}}\right),
\end{equation}
where we define
\begin{equation}
t_{\rm eff} \equiv \frac{m \gamma}{\mu \lambda} \left( 1- e^{-\mu \lambda
  t/[\gamma m]}
\right).
\end{equation}
Thus the distribution of the particle position in an harmonic potential
can be obtained from the distribution in the free L{\'e}vy flight case at
an earlier, ``effective'' time $t_{\mbox{eff}}$. This comparison
illustrates the slowing down of the particle in the harmonic potential,
where the restoring force is centered towards the origin. It
characterizes in a precise way the approach to stationarity, which
is graphed in Fig. \ref{fig6}. Furthermore, this approach is seen to take
longer time the smaller $\mu$ is; the quickest relaxation occurs in the
Brownian case $\mu=2$.
%\begin{figure}
%\unitlength=1cm
%\begin{center}
%\begin{picture}(6,6.2)
%\put(-2.2,6.6){
%\special{psfile=fig6.ps hoffset=0
%         voffset=0 hscale=34 vscale=34 angle=270}}
%\end{picture}
%\end{center}
%\caption{
%The linear time as seen by the free walker (dashed line), compared
%to the effective time sensed by the random walker in the harmonic
%potential, as a function of the laboratory time. For the L{\'e}vy
%flight in the potential, the restoring Hookean force slows down the
%spreading of the diffusion, and eventually brings it to a halt.
%\label{fig6}}
%\end{figure}

We conclude this section by some remarks about the inertial
term encountered for the Hookean force. There are two time scales (decay
times) involved in this case:
\begin{eqnarray}
\label{timescales}
\tau_s^{-1}&\equiv&
\frac{\gamma}{2}\left(1-\sqrt{1-\frac{4\lambda}{m\gamma^2}}\right)\\
\tau_f^{-1}&\equiv&\frac{\gamma}{2}\left(1+\sqrt{1-\frac{4\lambda}{
m\gamma^2}}\right)
\end{eqnarray}
Considering only the case of large over damping, i.e. $\gamma^2\gg
4\lambda/m$, the two time scales separate into a fast and a slow decaying
mode, $\tau_f\rightarrow \gamma^{-1}$ and $\tau_s\rightarrow \gamma
m/\lambda$, with $\tau_f\ll\tau_s$. It can be easily shown, that neglecting
the fast mode for times long compared to $\tau_f$, corresponds to
neglecting the inertial term in the original Langevin equation. Thus we
have an effective separation
into three different regimes, and being in the second with
$\tau_f\ll t<\tau_s$, the approach to the stationary state is not
influenced by the omission of the inertial term.

\section{Solution of the Langevin equation}
\label{langevin}

All of the above results could have been reached equally well
directly from the Langevin equation Eq. (\ref{langevin1}). To illustrate
this, we solve this equation for a L{\'e}vy flight in a constant
force field in addition to a linear force, $F(x)=-\lambda x+F_0$
corresponding for instance to a harmonic potential and a superimposed
gravity field. However, it could also correspond to many harmonic
oscillators placed at different positions, i.e.
\begin{eqnarray}
\nonumber
F(x)&=&\sum_{i=1}^N-\lambda_i(x-x_i)=-\left(\sum_{i=1}^N\lambda_i\right)x
+\sum_{i=1}^N\lambda_i x_i\\
&=&-\lambda x+F_0.
\label{harmpots}
\end{eqnarray}
The solution of Eq. (\ref{langevin1}) is given by
\begin{equation}
\label{langevin_sol}
x(t)=e^{-\lambda t/[\gamma m]}\int_0^t dt'e^{\lambda t'/[\gamma m]}
\left(\eta (t')+\frac{F_0}{\gamma m}\right).
\end{equation}
The distribution can then be found using the identity
\begin{eqnarray}
\nonumber
p(x,t)&=&\langle\delta (x-x(t))\rangle\\
\nonumber
&=&\int \frac{dk}{(2\pi)} \langle\exp(ik[x-x(t)])\rangle\\
&=&\int \frac{dk}{(2\pi)}e^{ikx}p(k,t).
\label{dist}
\end{eqnarray}
Using the solution Eq. (\ref{langevin_sol}) to obtain for the
characteristic function $p(k,t)$
\begin{mathletters}
\begin{eqnarray}
\nonumber
p(k,t)&=&\langle \exp\left(-i e^{-\lambda t/[\gamma m]}k\right.\\
&&\times \left. \int_0^t dt' e^{\lambda
t'/[\gamma m]}\left[\eta(t')+\frac{F_0}{\gamma m}\right]
\right)\rangle,
\label{discrete}
\end{eqnarray}
we have after discretizing the integral
\begin{eqnarray}
\nonumber
p(k,t)
& \simeq & e^{-ikF_0/\lambda (1-e^{-\lambda t/[\gamma m]})}\\
&&\times \langle
\prod_{t'=0}^t \exp\left(-ie^{-\lambda (t-t')/[\gamma m]}\Delta
k\eta(t')\right)\rangle
\end{eqnarray}
or
\begin{eqnarray}
\nonumber
p(k,t)
&\simeq&e^{-ikF_0/\lambda (1-e^{-\lambda t/[\gamma m]})}\\
&&\times\prod_{t'=0}^t
\langle \exp\left(-ie^{-\lambda (t-t')/[\gamma m]}\Delta
k\eta(t')\right)\rangle.
\end{eqnarray}
Using the definition of L{\'e}vy noise from Eq. (\ref{levy}), we
obtain
\begin{eqnarray}
\nonumber
p(k,t)&\simeq&e^{-ikF_0/\lambda (1-e^{-\lambda t/[\gamma m]})}\\
&&\times\prod_{t'=0}^t
\exp\left(-De^{-\mu\lambda (t-t')/[\gamma m]}\Delta^\mu |k|^\mu\right)
\label{expectation}
\end{eqnarray}
\end{mathletters}
Reintroducing the integrals and using the same renormalization
$D\Delta^{\mu-1}\rightarrow D$ as in passing from Eq. (\ref{langevin1}) to
Eq. (\ref{fp1}), we finally have
\begin{eqnarray}
\nonumber
p(k,t)&=& e^{-ikF_0/\lambda (1-e^{-\lambda t/[\gamma m]})}\\
&&\times \exp\left(-D\gamma m\left[1-e^{-\mu\lambda t/[\gamma m]}
\right]
\frac{|k|^\mu}{\mu\lambda}\right)
\end{eqnarray}
For $\lambda=0$, we recover the constant force result, Eq.
(\ref{sol2_fourier}). For $F_0=0$ we get Eq. (\ref{lin_four_sol}).
The free L{\'e}vy flight result $W_0(x,t)$ according to Eq.
(\ref{solution1}) is likewise reproduced, when $\lambda=F_0=0$.
In fact, by comparing to Eq. (\ref{solution1}) we have the
correspondence
\begin{eqnarray}
\nonumber
&&W_{\lambda,F_0}(x,t)=\\
&&W_0\left(x-\frac{F_0}{\lambda}\left[1-e^{-\lambda t/[\gamma m]}
\right],
\frac{\gamma m}{\mu\lambda} \left[1-e^{-\lambda t/[\gamma m]}
\right] \right).
\label{correspondence2}
\end{eqnarray}
In the presence of the harmonic potential we can no longer simply make a
Galilean transformation in order to eliminate the constant force, since the
presence of the linear force singles out another special reference frame.

\section{Some Remarks on the Numerical Simulations}
\label{simulations}

Using a computer code written in C, we have simulated L{\'e}vy flights in
two dimensions in order to compare with the theoretical predictions. The
noise has been defined by the asymptotics of the L\'evy distribution to
\begin{equation}
\label{num_noise}
p(\eta)=\frac{\mu\eta_0^\mu}{2}|\eta|^{-1-\mu}
\end{equation}
We have basically
investigated three properties: 1) histograms of the distribution of
position
for the free walker, 2) histograms of the walker in a harmonic potential,
and 3) the dynamic exponent as defined via the imaginary box in Eq.
(\ref{meansquare}).

The imaginary box according to Eq. (\ref{meansquare}) grows in time
like the characteristic width of the stable distribution, $\langle
x^2(t)\rangle_L\sim t^{2/\mu}$, which is {\it not\/} the variance.
It gives a measure, that a finite portion of the probability is
gathered within a given interval, which we call the imaginary box.
The values of $L_1$ and $L_2$ have been chosen so as to 1) ensure that we
are in the asymptotic regime, where $W(x,t)\sim t|x|^{-1-\mu}$, and 2) to
produce good statistics for all values of $\mu$. When fitting the results
to a straight line on a log--log plot, we have selected a subset of
equidistant points from the entire set of data, in order not to favor the
high $t$ region over the low $t$ region.

Concerning the histogram, several precautions have
to be taken when working with power law statistics. First of all, due to
the
occurrence of arbitrarily long steps, we have to define the interval of
sampling beforehand, since this is the only way to improve the statistics
when increasing the number of samples. We have chosen a minimum limit for
the evaluated data points to ensure the asymptotic range. The maximum limit
has been chosen as the maximum of the first say 100 (out of a total of
10000) simulations. In this way
we obtain many data points throughout the entire region, where the
asymptotic
power--law expression is valid.

We close with a remark on the axes of the plots of the histograms. To
obtain equidistant
points on the log--log plots, we chose to graph the histograms of the
distributions $p(y)$ of $y$, where $y=\log x$. If now $p(x)\sim
|x|^{-1-\mu}$, we have $p(y)\sim e^{-\mu y}$.
Plotting the logarithm of $p(y)$ as a function of $y$, we find a straight
line with slope $-\mu$, which is not to be confused with the slope
$-1-\mu$ for the log--log plot of $p(x)$.

\section{Conclusions}

We have investigated L{\'e}vy flights under the influence of external
force fields. Especially for the cases of free flights, a constant
and a linear force explicit solutions are derived and the consequences
shown. The solutions can be reduced to a transformation of variables
in the free flight result. We have employed the approaches of FFPEs
and a Langevin equation
with a power law noise term. We have shown that the
classical fluctuation--dissipation theorem is violated in the case of
constant force, and exhibited the non--Gibbsian nature of the
``equilibrium'' distribution in the harmonic potential. The approach to
stationarity has been characterized by an effective time, and the
connection to Tsallis's $q$--statistics has been explored in some detail.
For the constant force, due to the divergence of the mean square
displacement, the generalized Einstein relation breaks down, so that
the FFPE Eq. (\ref{fp1}) also violates linear response.
Numerical simulations have been found to agree well with the theoretical
analysis.

The direct solution of the Langevin equation Eq. (\ref{langevin1}) was
given, in agreement with the solutions for the FFPE approach. The
Langevin approach contains {\it a priori\/} more information than the
corresponding FFPE, and might be more suitable for a physical
interpretation of the underlying system. For a given problem, however, it
will be more convenient to employ the FFPE approach and use the methods of
characteristics or the separation of variables. Especially, the extraction
of moments $\langle x^n \rangle$ is straightforward using the FFPE, by
noting that $\frac{d}{dt}\langle x^n \rangle = \int dx \dot{W}(x,t)x^n$.

L{\'e}vy flights are typical for systems off thermal equilibrium, where
thermal equilibrium is to be understood in the classical Boltzmann--Gibbs
sense. For the systems analyzed in Refs. \cite{rocks,react,sms,albatros},
or
similar situations, the remoteness from thermal equilibrium is evident.
For the bulk mediated surface diffusion process in Ref.
\cite{bychuk,kimmich1}, one
should keep in mind, that the L{\'e}vy process emanates only as an
{\it effective\/} motion, resulting from a Brownian walker which is
eventually absorbed on the surface, before he gets activated again.

The connections to the Tsallis's $q$--entropy still remain unclear.
The discrepancy of the relations of $q$ to the L{\'e}vy index $\mu$
in the free and harmonic cases show, that the concept of $q$--entropy
cannot provide a full explanation of L{\'e}vy flights. Especially,
the stationary solution found for the harmonic potential, is not an
equilibrium solution in the sense of Tsallis's $q$--entropy. To our
understanding, only a non--linear generalized FPE can lead to results
compatible with this theory.

We believe that our analysis provides further understanding
of anomalous diffusion processes, and will give rise to further
experimental investigations, for example L{\'e}vy type reaction
dynamics subject to an electric field, or the tracer diffusion in
rock structures under gravitation, or similar.

\section{Acknowledgments}

We thank Yossi Klafter for helpful discussions. S.J. wishes to thank
Signe N\o rh\o j for useful comments. R.M. thanks the University of
{\AA}rhus for the invitation and the kind hospitality at the Institute
of Physics and Astronomy, and acknowledges financial support from the
Alexander von Humboldt Stiftung, Bonn am Rhein, Germany, through a
Feodor--Lynen fellowship, as well as from MINERVA through an Amos de
Shalit fellowship. Financial support from the German Israeli Foundation
(GIF) is acknowledged as well. R.M. thanks Israela Becker for
instigating discussions.

\begin{appendix}

\section{Correlation Functions}

Here we examine two--point correlations like $\langle
x(t)x(t')\rangle$ in the harmonic case, according to the tools developed in
Sec. \ref{langevin}. They are often sufficient for the description of
the system in applications of the underlying (F)FPE. Assuming the
initial condition $x(0)=0$, we introduce the Green's function
\begin{equation}
x(t)=\int dt' \, G(t,t')\eta(t'),
\end{equation}
and by comparison with Eq. (\ref{langevin_sol}) it is seen that
\begin{equation}
G(t,t')=e^{-\lambda (t-t')/\gamma m}\Theta (t-t')\Theta (t').
\end{equation}
Due to the divergence of the moments of the L\'evy distribution
in general, we have to work with characteristic functions discussing
correlations:
\begin{eqnarray}
\nonumber
&&\hspace{-0.4in}\langle e^{-i\int ds \, A(s)x(s)}\rangle\\
&=&\langle\exp\left(-i\int
ds\int ds'\,A(s)G(s,s')\eta(s')\right)\rangle\nonumber\\
&\simeq &\prod_{s'}\langle e^{-i\Delta\eta(s')\int ds \,
A(s)G(s,s')}\rangle\nonumber\\
&=&\prod_{s'}e^{-D\left|\Delta\int ds\, A(s)G(s,s')\right|^{\mu}}
\end{eqnarray}
where $A(t)$ denotes an a priori arbitrary function, and we used the
definition Eq. (\ref{levy}). With the usual renormalization of the noise
$D\Delta^{\mu-1}\rightarrow D$ \cite{fogedby1} we have
\begin{eqnarray}
\nonumber
&&\hspace{-0.2in}\langle e^{-i\int ds\, A(s)x(s)}\rangle\\
&=&\exp\left( -D\int ds'\left|\int ds\, A(s)G(s,s')\right|^{\mu}
\right)\nonumber\\
\nonumber
&=&\exp\left(-D\int ds'\right.\\
&& \times \left. \left|\int ds \int ds'' G(s,s')G(s'',s')
A(s)A(s'')\right|^{\mu/2}\right)\nonumber\\
\nonumber
&=&\exp\left(-D\int_0^\infty e^{\mu\lambda s'/[\gamma m]} ds'
\left| \int_{s'}^\infty ds \right. \right.\\
&& \times \left. \left.\int_{s'}^\infty ds''
e^{-\lambda (s+s'')/[\gamma m]}A(s) A(s'')\right|^{\mu/2}\right).
\label{generator}
\end{eqnarray}
In the last step we used the identity
\begin{eqnarray}
\nonumber
G(s,s')G(s'',s')&=&e^{-\lambda(s+s''-2s')/[\gamma m]}\\
&& \times \Theta(s-s')\Theta(s''-s')\Theta(s') .
\end{eqnarray}
Putting $A(s)=A\delta(t-s)$, we find
\begin{eqnarray}
\label{onepoint}
&&\hspace{-0.2in}\langle e^{-iA x(t)}\rangle\nonumber\\
\nonumber
&=&\exp\left(-D \int_0^\infty ds' e^{\mu\lambda s'/[\gamma m]}
\left|\int_{s'}ds \right. \right.\\
&& \times \left. \left. \int_{s'}ds'' e^{-\lambda
(s+s'')/[\gamma m]}A^2\delta(s-t)\delta(s''-t)\right|^{\mu/2}
\right)\nonumber\\
\nonumber
&=&\exp\left(-D\int_0^\infty ds' \right.\\
&&\times \left. e^{\mu\lambda s'/[\gamma m]}\left|
e^{-2\lambda t/[\gamma m]}A^2\theta(t-s')\right|
^{\mu/2}\right)\nonumber\\
&=&\exp\left(-D\int_0^{t}dse^{-\mu\lambda(t-s)/[\gamma
m]}A^{\mu}\right)\nonumber\\
&=&\exp\left(-DA^\mu\left[\frac{\gamma m(1-e^{-\mu\lambda
t/[\gamma m]})} {\mu\lambda}\right]\right),
\end{eqnarray}
which is equivalent to Eq. (\ref{lin_four_sol}), as it should be, and
therefore
includes
also the results for the Brownian case. One--point correlation functions
(moments of the distribution), if they exist, can thus be obtained from
Eq. (\ref{onepoint}), so we will proceed to the more interesting case
of the two--point correlations. To this end, we take $A(t)= A\left[
\delta(t-t_1)-\delta (t-t_2)\right]$ and insert it into Eq.
(\ref{generator}),
and we find
\begin{eqnarray}
&&\hspace{-0.4in}\langle
\exp\left(-iA(x(t_1)-x(t_2))\right)\rangle=\nonumber\\
\nonumber
&&\exp\left(-DA^\mu \left[ \frac{\gamma m(1-e^{-\mu\lambda
t_2/[\gamma m]})} {\mu\lambda} \right. \right.\\
\nonumber
&&\hspace{0.4in}\times \left. \left. \left(1-e^{-\mu\lambda
(t_1-t_2)/[\gamma m]}\right)^\mu \right. \right.\\
&& \hspace{0.4in}\left. \left. +\frac{\gamma
m(1-e^{-\mu\lambda(t_1-t_2)\gamma m})} {\mu\lambda}\right]\right)
\label{twopoint}
\end{eqnarray}
for $t_1>t_2$, and $t_1$ and $t_2$ interchanged in the other case
when $t_2>t_1$. This is essentially the characteristic function of the
stochastic variable $x(t_1)-x(t_2)$, so all two--point correlation
functions and the distribution itself can in principle be found
from Eq. (\ref{twopoint}). However, it is seen that it is a L{\'e}vy
distribution, so all higher moments diverge. Nevertheless, Eq.
(\ref{twopoint}) still gives some information about the correlation
between the position of
the walker at two different times.  When $t_1\gg t_2$ the characteristic
function splits up into the product of the characteristic function of the
two variables Eq. (\ref{onepoint}) which means, that $x(t_1)$ and $x(t_2)$
are independent in this limit. At intermediate times, i.e.
when both $t_1$ and $t_2$ are small, the correlation depends on
both. This is a memory of the initial conditions, since both walkers
start out at the origin. At long times, initial conditions are not
important, and hence only a dependence on the time difference $t_1-t_2$
(kept fixed and finite) is retained,
\begin{eqnarray}
\nonumber
&&\hspace{-0.4in}\langle e^{-iA(x(t_1)-x(t_2))}\rangle\nonumber\\
\nonumber
&&=\exp\left(\frac{-D\gamma m}{\mu\lambda}
\left[\left(1-e^{-\mu\lambda|t_1-t_2|/[\gamma
m]}\right)^\mu \right. \right.\\
&& \hspace{0.4in}\left. \left. +1-e^{-\mu\lambda|t_1-t_2|/[\gamma m]}
\right] \right).
\label{longtimes}
\end{eqnarray}
Writing $x_{12}\equiv x(t_1)-x(t_2)$, we have by the usual arguments a
L\'evy
distribution of $x_{12}$:
\begin{equation}
\langle e^{-i k\cdot x_{12}}\rangle
=e^{-\tilde{D}_{\mu}(t_1,t_2)|k|^\mu}
\end{equation}
with
\begin{eqnarray}
\nonumber
&&\tilde{D}_{\mu}(t_1,t_2)\equiv D\frac{\gamma m}{\mu \lambda}\left[
1-e^{-\mu\lambda t_2/[\gamma m] } \left(1-e^{-\mu\lambda
(t_1-t_2)/[\gamma m]}\right)^\mu \right.\\
&& \hspace{0.4in}\left. +1-e^{-\mu\lambda(t_1-t_2)/[\gamma m]}\right].
\label{x12}
\end{eqnarray}
As with the other L\'evy distributions, this stochastic variable is
characterized by a power law tail ($t_1\equiv t$, $t_2\equiv 0$):
\begin{eqnarray}
\nonumber
p(x_{12},t)&\sim&
D\frac{\gamma m}{\mu\lambda}\left((1-e^{-\mu\lambda t/[\gamma
m]})^\mu \right.\\
&&\hspace{0.2in}\left.
+1-e^{\mu\lambda t/[\gamma m]}\right)|x_{12}|^{-1-\mu}.
\label{tail2}
\end{eqnarray}
In the case of $\mu=2$ all the preceding results reproduce the
well--known Brownian relations for ordinary diffusion. One could
proceed like this finding three--point correlations in an analogous
manner.

\end{appendix}

\clearpage

\begin{thebibliography}{99}
\bibitem{rem2} A renormalisation involving ``$dt$'' and $D$ has to be
carried out in order to make the problem well--defined
\protect\cite{fogedby}.
\bibitem{rem1} This asymptotics could also be taken as the definition of
the noise, with a cutoff $\eta_0$ to account for normalisability
\protect\cite{fogedby1,fogedby}.
\bibitem{comment} The scaling result $\langle x^2 (t) \rangle_L \sim t^{
2/\mu}$ [Eq. (\protect\ref{meansquare})] and the squared absolute mean
$\langle |x|\rangle ^2$
[Eq. (\protect\ref{sam})] are proportional to $t^{2/\mu}$. The divergence
of the mean square displacement $\langle x^2 \rangle=\int dx \, x^2 W(x,t)$
can be avoided in the L{\'e}vy walk concept through the introduction of
a time cost which penalizes long displacements. For L{\'e}vy walks,
the proportionality $\langle x^2 \rangle=\int dx W(x,t)x^2\propto t^{3-
\gamma}$ is found, see \protect\cite{klablushle} and \protect\cite{reply}.
\end{thebibliography}
\end{document}